\documentclass[a4paper,12pt]{article}
\usepackage{geometry}

\usepackage{hyperref}
\usepackage{xcolor}
\hypersetup{
colorlinks,
linkcolor={blue},
citecolor={blue},
urlcolor={blue!80!black}
}

\usepackage{upgreek}

\usepackage{graphicx}
\usepackage{epstopdf}
\usepackage{mathrsfs}
\usepackage{textcomp}
\usepackage{stmaryrd} 

\usepackage{authblk}
\usepackage{graphicx}
\usepackage{epstopdf}
\usepackage{mathrsfs}
\usepackage{textcomp}
\usepackage{mathtools}

\usepackage{amssymb,amsmath,amsthm}

\newtheorem{remark}{Remark}

\newtheorem{theorem}{Theorem}

\usepackage{cancel}

\DeclareMathOperator{\tr}{tr}
\newcommand{\mc}[1]{\mathcal{#1}}
\newcommand{\mbs}[1]{\boldsymbol{#1}}
\newcommand{\mbf}[1]{\mathbf{#1}}

\newcommand{\mbb}[1]{\mathbb{#1}}
\newcommand{\wt}[1]{\widetilde{#1}}
\newcommand{\wh}[1]{\widehat{#1}}
\newcommand{\ol}[1]{\overline{#1}}
\newcommand{\pd}[2]{\frac{\partial#1}{\partial#2}}

\newcommand{\sfn}[1]{\mathscr{#1}}
\newcommand{\mbsi}[1]{\mathrm{#1}}

\usepackage{siunitx}
\numberwithin{equation}{section}

\usepackage[symbol]{footmisc}
\makeatletter
\def\blfootnote{\xdef\@thefnmark{}\@footnotetext}
\makeatother

\usepackage{array}
\newcolumntype{L}[1]{>{\raggedright\let\newline\\\arraybackslash\hspace{0pt}}m{#1}}
\newcolumntype{C}[1]{>{\centering\let\newline\\\arraybackslash\hspace{0pt}}m{#1}}
\newcolumntype{R}[1]{>{\raggedleft\let\newline\\\arraybackslash\hspace{0pt}}m{#1}}

\usepackage{tabu}

\title{Null Lagrangians in Cosserat elasticity}

\author{Basant Lal Sharma$^1$\thanks{Corresponding author email: bls@iitk.ac.in}, Nirupam Basak$^2$, \\
{\small $^1$Department of Mechanical Engineering, Indian Institute of Technology Kanpur}\\
{\small Kanpur, U. P. 208016, India; email{bls@iitk.ac.in}}\\
{\small $^2$Department of Mathematics, Indian Institute of Technology Kanpur}\\
{\small Kanpur, U. P. 208016, India; basakn@iitk.ac.in}}

\date{}

\allowdisplaybreaks 

\begin{document}

\maketitle

\begin{abstract}
In the framework of nonlinear theory of Cosserat elasticity, also called micropolar elasticity, we provide the complete characterization of null Lagrangians for three dimensional bodies as well as for shells. Using the Gibb's rotation vector for description of the microrotation, this task is possible by an application of a theorem stated by Olver and Sivaloganathan in `{the structure of null Lagrangians}' (Nonlinearity, {1}, 1988, pp. 389-398).
A set of necessary and sufficient conditions is also provided for the elasticity tensors to correspond to a null Lagrangian in linearized micropolar theory.

\end{abstract}

\section*{Introduction}

According to Theorem 7 of Olver and Sivaloganathan in \cite{OS88},
for a star shaped ${{\si{\ohm}}}\subset\mbb{R}^m$, with
\[
\mbs{x}\in{{\si{\ohm}}}, \qquad\mbs{u}:\ol{{{\si{\ohm}}}}\to\mbb{R}^n,\qquad \mbf{F}:{{\si{\ohm}}}\to\mbb{M}^{n\times m},
\]
a function $L(\mbs{x},\mbs{u},\mbf{F})$ (with $\mbf{F}=\nabla \mbs{u}=[{\partial u_i}/{\partial x_j}]\in\mbb{M}^{n\times m}$) is a {\em null Lagrangian} if and only if there exist 
an $m$-tuple of $C^1$ functions 
\[
\mbs{P}:\ol{{{\si{\ohm}}}}\times\mbb{R}^n\times\mbb{M}^{n\times m}\to\mbb{R}^m
\]
such that
\[
\sfn{L}(\mbs{x},\mbs{u},\nabla \mbs{u})={\nabla\cdot} \mbs{P}(\mbs{x},\mbs{u},\nabla \mbs{u})~\forall \mbs{u}\in C^1({{\si{\ohm}}}),
\]
where the arbitrary scalar potential functions on $\ol{{{\si{\ohm}}}}\times\mbb{R}^n$ participating in the divergence above via the presence of $\mbs{P}$ can be also specified while their total number is given by the binomial coefficient $\binom{n+m}{m-1}$.

In the familiar case of three dimensional theory of elasticity, the number of arbitrary scalar potentials is known to be $\binom{3+3}{3-1}$ (with $m=3, n=3$), i.e., $15$.
For our purpose in this note, as another example, in the case of three dimensional {\em Cosserat} (micropolar) theory \cite{Cosserat09}, the number of arbitrary scalar potentials in the sum appearing in Theorem 8 of Olver and Sivaloganathan \cite{OS88} is anticipated to be $\binom{3+3+3}{3-1}$ (with $m=3, n=3+3=6$), i.e., 36,
whereas for two dimensional shell theory (embedded in three dimensional space), this number is $\binom{2+3+3}{2-1}$, i.e., 8 (with $m=2, n=3+3=6$).

For the benefit of some readers, we recall that a null Lagrangian (see \cite{Ball81}, \cite{AD80}, \cite{OS88}, \cite{Olver83}; \cite{dE62}, \cite{Rund66}, \cite{dE86}, \cite{Landers42}, \cite{Ericksen62}) is defined by the condition that its Euler--Lagrange equation is trivially satisfied; in other words, the so called functional $\mc{L}$ given by the expression
\[
\mc{L}[\mbs{u}]=\int_{{{\si{\ohm}}}}\sfn{L}(\mbs{x},\mbs{u}(\mbs{x}),\nabla \mbs{u}(\mbs{x}))d{x}
\]
satisfies 
\[
\mc{L}[\mbs{u}+\mbs{\varphi}]=\mc{L}[\mbs{u}]~\forall \mbs{\varphi}\in C^\infty_0({{\si{\ohm}}}).
\]
In the nonlinear theory of elasticity, the null Lagrangians have been found to have special importance in the questions of existence of solutions \cite{Ball77,Ball81,Dac2010,Tadeusz1993} as well as in the surface potentials and handling certain boundary data \cite{Carillo02,dE81,dE86}. The connection with the construction of polyconvex functions has a practical value as it is helpful in developing the riogorous framework for some very useful elastic models \cite{Ogden97,Steigmann2003} and also in the presence of various additional physical effects \cite{Itskov2016,Silhavy2018}. Besides this the role of null Lagrangians in Noether symmetries is also well known \cite{Noether18,BH21,NZ98}.
Last but not the least, in the classical framework of calculus of variations \cite{GH04}, the null Lagrangians occupy a distinguished role in the field theory as any researcher can easily find out during an expedition on the `royal road of Caratheodory'.

In this short note, we apply the above mentioned Theorem of \cite{OS88} to a Cosserat \cite{Cosserat09}, or so called, \emph{micropolar} elastic body \cite{Aero1,Aero2,Eringen99,wN86}.

\section{Nonlinear Cosserat media}
\begin{figure}\centering\includegraphics[width=.7\linewidth]{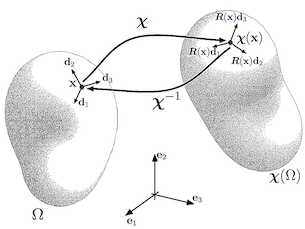}
\caption{Kinematics for a Cosserat (micropolar) body.}
\label{micropolarbody}\end{figure}

We consider a body of Cosserat type with the reference configuration assumed to be a bounded domain denoted by ${{\si{\ohm}}}\subset{\mathbb{R}}^3$ (with a Lipschitz boundary $\partial{{\si{\ohm}}}$).
However, it suffices to consider any smooth portion of the body as we are interested only in the characterization of the null Lagrangians on the lines of that in the nonlinear theory of elasticity \cite{Silhavy97}. 
Following the standard notation for vectors and tensors in continuum mechanics \cite{Gurtin1981}, we denote the \emph{microdeformation} (vector field), or placement, of a micropolar body by
\begin{equation}
\label{deform}
{\mbs{\chi}}:{{\si{\ohm}}}\rightarrow\mathbb{R}^3
\end{equation}
and the \emph{microrotation} (describing the rotation of each particle in the micropolar body) with
\begin{equation}
\label{deformR}
{\mbf{R}}:{{\si{\ohm}}}\rightarrow\mathrm{SO(3)}.
\end{equation}
An schematic is provided in Fig. \ref{micropolarbody} which illustrates the manner in which the microrotation field captures the rotation of an orthonormal triad of directors from reference configuration ${\si{\ohm}}$ to the current configuration ${\mbs{\chi}}({\si{\ohm}})$.

Here, we denote the standard basis vectors for $\mathbb{R}^3$ by the triplet $\mbf{e}_1, \mbf{e}_2, \mbf{e}_3$. The physical space $\mathbb{R}^3$ is assumed to be equipped with the cross product $\times$ corresponding to an orientation such that $\mbf{e}_1\times\mbf{e}_2\cdot \mbf{e}_3=+1$. In \eqref{deformR}, we use the symbol $\mathrm{SO(3)}$ to denote the set of all rotation tensors in three dimensions, i.e., for $\mbf{Q}\in\mathrm{SO(3)}$, $\mbf{Q}^{\top}\mbf{Q}=\mbf{I}, \det \mbf{Q}=+1$, where $\mbf{I}$ stands for the identity tensor and ${}^{\top}$ denotes the transpose.
For a skew tensor $\mbf{W}$ (i.e., $\mbf{W}^{\top}=-\mbf{W}$) the axial vector
$\mbf{w}=\mathrm{axl}\mbf{W},$
is defined by
$\mbf{W}\mbf{a}=\mbf{w}{\times}\mbf{a}, \forall \mbf{a}\in\mathbb{R}^3.$
The 
relation $\mbf{W}=-\mbs{\epsilon}\mbs{w}$ provides the skew tensor corresponding to a given vector, where $\mbs{\epsilon}$ is the three dimensional alternating tensor which plays the role of a linear map from vectors $\mathbb{R}^3$ to tensors $\mathbb{M}^{3\times3}$ here; in components, the skew tensor $\mbf{W}$ corresponding to a vector $\mbs{w}$ is given by
$\mbsi{W}_{ij}=-\upepsilon_{ijk}\mbsi{w}_k$ with $\upepsilon_{ijk}=\mbs{e}_i\cdot \mbs{e}_j\times\mbs{e}_k$.
In other words,
$\text{skew}(\mbs{w}):=-\mbs{\epsilon}\mbs{w}.$
We also employ 
a very convenient notation \cite{Silhavy97} for a related entity, also called vector invariant (or Gibbsian Cross), $\mbf{A}^{\times}$ with the components
\begin{equation}
(\mbf{A}^{\times})_i := \upepsilon_{ijk} \mbsi{A}_{jk}
\label{crossM}
\end{equation}
for any second order tensor $\mbf{A}$. Thus, $\mathrm{axl}\mbf{W}=-\frac{1}{2}\mbf{W}^{\times}$ for skew tensor $\mbf{W}$.
The differentiation of a function $f$ (which depends on position vector $\mbs{x}$) with respect to the $x_j$ coordinate is written as $f_{,j}.$ Note that $\mbf{Q}^{\top}\mbf{Q}_{,j}$ is a skew tensor for a rotation tensor field $\mbf{Q}$ on ${{\si{\ohm}}}$.

With above notation in place, the deformation gradient corresponding to \eqref{deform} is expressed as
\begin{equation}
\mbf{F}:=\nabla{\mbs{\chi}}= {\mbs{\chi}}_{,i} \otimes \mbf{e}_i, \qquad\forall \mbf{x}\in{{\si{\ohm}}},
\end{equation}
while the nonsymmetric right stretch tensor is defined as
\begin{equation}
\mbf{U}:={\mbf{R}}^{\top}\mbf{F}.
\end{equation}
We define the relative Lagrangian {\em stretch tensor} as strain measure by \cite{PE09}
\begin{equation}
{\mbf{E}}:={\mbf{U}}- {\mbf{I}}.
\end{equation}
In micropolar media, an additional dependent field is the axial vector of ${\mbf{R}}^{\top}{\mbf{R}}_{,j}$.
The second order tensor
\begin{equation}
{\mbf{K}}:= \mathrm{axl}\big({\mbf{R}}^{\top} {\mbf{R}}_{,j} \big)\otimes {\mbf{e}}_j,
\end{equation}
is a Lagrangian measure for curvature \cite{PE09}, called the {\em wryness tensor}. 
In the nonlinear theory of Cosserat, i.e., the micropolar elasticity, the strain energy density function (in terms of the tensors of stretch ${\mbf{E}}$ and wryness ${\mbf{K}}$) is
\begin{equation}
\label{WEK}
W({\mbf{E}},{\mbf{K}}).
\end{equation}
In order to proceed further for the characterization of the null Lagrangians, it is useful to employ the local coordinates for the 
microrotation ${\mbf{R}}$ \eqref{deformR}. 
We utilize the Gibb's rotation vector (or coordinates) to express the rotation $\mbf{R},$
\begin{equation}
\label{Gibbsrot}
{\mbf{R}}={\mbf{R}}({\boldsymbol{\theta}}):=\frac{1}{4+\theta^2}((4-\theta^2)\mbf{I}+2{\boldsymbol{\theta}}\otimes{\boldsymbol{\theta}}-4~\text{skew}(\mbs{\theta})), \qquad\theta^2={\boldsymbol{\theta}}\cdot{\boldsymbol{\theta}}.
\end{equation}
It is easy to show that
\[
\text{axial}({\mbf{R}}^{\top}{\mbf{R}}_{,i})=-\frac{1}{4+\theta^2}(4{{\boldsymbol{\theta}}}_{,i}+2{\boldsymbol{\theta}}\times{{\boldsymbol{\theta}}}_{,i}),
\]
so that
\[
\mbf{K}=-\frac{1}{4+\theta^2}(4\nabla{\boldsymbol{\theta}}+2{\boldsymbol{\theta}}\times{{\boldsymbol{\theta}}}_{,i}\otimes\mbs{e}_i).
\]
In the context of the energy functionals (based on \eqref{WEK}, for example) for a micropolar elastic body,
we consider the null Lagrangian for the corresponding class of functionals
\begin{equation}
\mc{L}[{\mbs{\chi}},{{\boldsymbol{\theta}}}]=\int_{{\si{\ohm}}} \sfn{L}(\mbs{x},{\mbs{\chi}},{\boldsymbol{\theta}},\nabla{\mbs{\chi}},\nabla{\boldsymbol{\theta}})d{x}.
\label{microlag}
\end{equation}

\begin{remark}\it
It is possible to combine ${\mbs{\chi}}, {\boldsymbol{\theta}}$ together as a single vector field $\mbs{u}$ in $\mathbb{R}^6$ however we refrain from doing this in the first and second section. We insist on retaining the original fields so that the analysis yields a decomposition of the terms which can be utilized directly by the reader in various applications of interest.
\end{remark}

\section{Null Lagrangian in three dimensional Cosserat theory}
\label{section3d}
Due to the presence of three different vector entities namely, $\mbs{x}, {\mbs{\chi}},$ and ${\boldsymbol{\theta}}$, it is convenient to employ a more delicate indicial notation. 
Henceforth, let the local coordinates be
denoted by $x_A$ for $\mbs{x}$ and 
$y_i$ for ${\mbs{\chi}}$. 
As explained above, the local coordinates for ${\boldsymbol{\theta}}$, essentially for ${\mbf{R}}$, are $\theta_\alpha$. 
In the assumed framework for three dimensional Cosserat body, we have the following identification of the local coordinates with components
\[
\mbs{x}=x_A\mbs{e}_A,\qquad
\mbs{y}={\mbs{\chi}}(\mbs{x})=y_i\mbs{e}_i,\qquad
{\boldsymbol{\theta}}=\theta_\alpha\mbs{e}_\alpha,
\]
where except for the indices the orthonormal triad of vectors $\{\mbs{e}_1, \mbs{e}_2, \mbs{e}_3\}$ can be chosen to be the same.
\begin{remark}\it
In indicial notation, according to \eqref{Gibbsrot},
\[
\mbsi{R}_{\alpha\beta}=\frac{1}{4+\theta_\eta\theta_\eta}((4-\theta_\eta\theta_\eta)\delta_{\alpha\beta}+2\theta_\alpha\otimes \theta_\beta+4\upepsilon_{\alpha\beta\gamma}\theta_\gamma),
\]
while the inverse relation can be easily found to be
\[
\theta_\alpha=\frac{2}{1+\mbsi{R}_{\eta\eta}}\upepsilon_{\alpha\beta\gamma}\mbsi{R}_{\beta\gamma},
\]
provided $\mbsi{R}_{\eta\eta}\ne-1$.
With $\mbs{\epsilon}$ in the role of a linear map from second order tensors to vectors, this can be also expressed as $\mbs{\theta}=({2}/({1+\tr{\mbf{R}}}))\mbs{\epsilon}\mbf{R}$.
\end{remark}

The following is based on the result of \cite{OS88,Gr99} for null Lagrangians (also termed as variationally trivial Lagrangians).
Let
\begin{equation}\begin{aligned}
\omega
&= \sfn{A} dx_1 {{\wedge}} dx_2 {{\wedge}} dx_3+\sfn{D}dy_1 {{\wedge}} dy_2 {{\wedge}} dy_3\\
&+\frac{1}{2}\upepsilon_{ABC}\mbsi{B}_{Ai}dy_i {{\wedge}} dx_B {{\wedge}} dx_C+\frac{1}{2}\upepsilon_{ijk}\mbsi{C}_{iA}dy_j{{\wedge}} dy_k {{\wedge}} dx_A \\
&+\frac{1}{2}\upepsilon_{ABC}\wt{\mbsi{B}}_{A\alpha}d\theta_\alpha {{\wedge}} dx_B {{\wedge}} dx_C 
+\frac{1}{2}\upepsilon_{i\beta\gamma}\wt{\mbsi{C}}_{iA}d\theta_\beta{{\wedge}} d\theta_\gamma {{\wedge}} dx_A+\wt{\sfn{D}}d\theta_1 {{\wedge}} d\theta_2 {{\wedge}} d\theta_3\\
&+\frac{1}{2}\upepsilon_{\alpha\beta \gamma}\wh{\mbsi{B}}_{\alpha i}dy_i {{\wedge}} d\theta_\beta {{\wedge}} d\theta_\gamma+\frac{1}{2}\upepsilon_{ijk}\wh{\mbsi{C}}_{i\alpha}dy_j{{\wedge}} dy_k {{\wedge}} d\theta_\alpha+J_{\alpha j C}d\theta_\alpha {{\wedge}} dy_j {{\wedge}} dx_C,
\label{omegafn2form}
\end{aligned}\end{equation}
which involves 
a total $84$ arbitrary functions (as expected this number equals $\binom{3+3+3}{3}$) of $\mbs{x}, {\mbs{\chi}},$ and ${\boldsymbol{\theta}}$.
With details provided in Appendix \ref{appA}, the null Lagrangians are described by the general expression:
\begin{equation}\begin{aligned}
&\sfn{A}+\mbf{B}^{\top} \cdot {\mbf{F}}+\mbf{C}\cdot{{\mathrm{cof~}}} {\mbf{F}}+\sfn{D}\det {\mbf{F}}+\wt{\mbf{B}}^{\top}\cdot {\mbf{G}}+\wt{\mbf{C}}\cdot{{\mathrm{cof~}}} {\mbf{G}}+\wt{\sfn{D}}\det {\mbf{G}}\\
&+\wh{\mbf{B}}\cdot ({{\mathrm{cof~}}}\mbf{G})\mbf{F}^{\top}+\wh{\mbf{C}}\cdot({{\mathrm{cof~}}}{\mbf{F}}){\mbf{G}}^{\top}+\mathtt{J}\cdot\mbs{e}_\alpha\otimes \mbs{e}_j \otimes(\mbf{G}^{\top}\mbs{e}_\alpha{\wedge}\mbf{F}^{\top}\mbs{e}_j),
\label{omegafn}
\end{aligned}\end{equation}
where
\begin{equation}\begin{aligned}
{\mbf{F}}=F_{iA}\mbs{e}_i\otimes\mbs{e}_A,
{\mbf{G}}=G_{\alpha A}\mbs{e}_\alpha\otimes\mbs{e}_A,\\
\mbf{B}=\mbsi{B}_{iA}\mbs{e}_i\otimes\mbs{e}_A,\qquad
\wt{\mbf{B}}=\wt{\mbsi{B}}_{\alpha A}\mbs{e}_\alpha\otimes\mbs{e}_A,\qquad
\wh{\mbf{B}}=\wh{\mbsi{B}}_{i\alpha}\mbs{e}_i\otimes\mbs{e}_\alpha,\\
\mbf{C}=\mbsi{C}_{iA}\mbs{e}_i\otimes\mbs{e}_A,\qquad
\wt{\mbf{C}}=\wt{\mbsi{C}}_{\alpha A}\mbs{e}_\alpha\otimes\mbs{e}_A,\qquad
\wh{\mbf{C}}=\wh{\mbsi{C}}_{i\alpha}\mbs{e}_i\otimes\mbs{e}_\alpha,\\
\mathtt{J}=J_{\alpha j C}\mbs{e}_\alpha\otimes\mbs{e}_j\otimes\mbs{e}_C.
\end{aligned}\end{equation}
We seek to obtain necessary and sufficient conditions on the coefficients in \eqref{omegafn} so that it prescribes any arbitrary null Lagrangian (given the hypothesis on ${\si{\ohm}}$ for the applicability of Poincar\'{e} Lemman \cite{OS88,Gr99}).
Indeed, the general form of the null Lagrangian of the form \eqref{microlag} is obtained by the exterior derivative of the $2$-form
\begin{equation}\begin{aligned}
\zeta
&=\frac{1}{2}\upepsilon_{ABC}\mbsi{L}_A dx_B{{\wedge}} dx_C+\mbsi{K}_{iA}dy_i {{\wedge}} dx_A+\frac{1}{2}\upepsilon_{ijk}\mbsi{M}_i dy_j {{\wedge}} dy_k\\
&+\wt{\mbsi{K}}_{\alpha A}d\theta_\alpha{{\wedge}} dx_A
+\frac{1}{2}\upepsilon_{\alpha\beta\gamma}\wt{\mbsi{M}}_\alpha d\theta_\beta{{\wedge}} d\theta_\gamma
+\mbsi{H}_{\alpha j}d\theta_\alpha{{\wedge}} dy_j,
\label{zeta3d}
\end{aligned}\end{equation}
where the coefficients are functions of $x_A, y_i, \theta_\alpha$,
which form a total number of 36 functions of $x_A, y_i, \theta_\alpha$. 
With details provided in Appendix \ref{appB},
we find that the characterizing condition $\omega=d\zeta$ (and the Poincar\'e Lemma \cite{OS88}) implies
\begin{equation}\begin{aligned}
\sfn{A}=\mbsi{L}_{A,A}, \qquad\sfn{D}=\mbsi{M}_{i,i},\qquad \wt{\sfn{D}}=\wt{\mbsi{M}}_{\alpha,\alpha},\qquad J_{\alpha j C}=(\mbsi{K}_{jC,\alpha}-\wt{\mbsi{K}}_{\alpha C,j}+\mbsi{H}_{\alpha j,C}),\\
\frac{1}{2}\upepsilon_{ABC}\mbsi{B}_{Ai}=(\frac{1}{2}\upepsilon_{ABC}\mbsi{L}_{A,i}- \mbsi{K}_{iC,B}), \qquad
\frac{1}{2}\upepsilon_{ijk}\mbsi{C}_{iA}=(\mbsi{K}_{kA,j}+\frac{1}{2}\upepsilon_{ijk}\mbsi{M}_{i,A}),\\
\frac{1}{2}\upepsilon_{ABC}\wt{\mbsi{B}}_{A\alpha}=(\frac{1}{2}\upepsilon_{ABC}\mbsi{L}_{A,\alpha}-\wt{\mbsi{K}}_{\alpha C,B}), \qquad
\frac{1}{2}\upepsilon_{i\beta\gamma}\wt{\mbsi{C}}_{iA}=(\wt{\mbsi{K}}_{\gamma A,\beta}+\frac{1}{2}\upepsilon_{\alpha\beta\gamma}\wt{\mbsi{M}}_{\alpha,A}),\\
\frac{1}{2}\upepsilon_{\alpha\beta \gamma}\wh{\mbsi{B}}_{\alpha i}=(\frac{1}{2}\upepsilon_{\alpha\beta\gamma}\wt{\mbsi{M}}_{\alpha,i}+\mbsi{H}_{\gamma i,\beta}), \qquad
\frac{1}{2}\upepsilon_{ijk}\wh{\mbsi{C}}_{i\alpha}=(\frac{1}{2}\upepsilon_{ijk}\mbsi{M}_{i,\alpha}+\mbsi{H}_{\alpha j,k}).
\notag
\end{aligned}\end{equation}
Using the properties of the alternative tensor, moreover,
starting from the second line above, the relations can be simplified as
\begin{subequations}
\begin{align}
\mbsi{B}_{Ai}&=\upepsilon_{ABC}(\frac{1}{2}\upepsilon_{PBC}\mbsi{L}_{P,i}- \mbsi{K}_{iC,B})=\mbsi{L}_{A,i}-\upepsilon_{ABC}\mbsi{K}_{iC,B},\\
\mbsi{C}_{iA}&=\upepsilon_{ijk}(\mbsi{K}_{kA,j}+\frac{1}{2}\upepsilon_{pjk}\mbsi{M}_{p,A})=\upepsilon_{ijk} \mbsi{K}_{kA,j}+\mbsi{M}_{i,A},\\
\wt{\mbsi{B}}_{A\alpha}&=\upepsilon_{ABC}(\frac{1}{2}\upepsilon_{PBC}\mbsi{L}_{P,\alpha}-\wt{\mbsi{K}}_{\alpha C,B})=\mbsi{L}_{A,\alpha}-\upepsilon_{ABC}\wt{\mbsi{K}}_{\alpha C,B},\\
\wt{\mbsi{C}}_{iA}&=\upepsilon_{i\beta\gamma}(\wt{\mbsi{K}}_{\gamma A,\beta}+\frac{1}{2}\upepsilon_{\alpha\beta\gamma}\wt{\mbsi{M}}_{\alpha,A})=\upepsilon_{i\beta\gamma}\wt{\mbsi{K}}_{\gamma A,\beta}+\wt{\mbsi{M}}_{i,A},\\
\wh{\mbsi{B}}_{\alpha i}&=\upepsilon_{\alpha\beta \gamma}(\frac{1}{2}\upepsilon_{\delta\beta\gamma}\wt{\mbsi{M}}_{\delta,i}+\mbsi{H}_{\gamma i,\beta})=\wt{\mbsi{M}}_{A,i}+\upepsilon_{A\beta \gamma}\mbsi{H}_{\gamma i,\beta},\\
\wh{\mbsi{C}}_{i\alpha}&=\mbsi{M}_{i,\alpha}+\upepsilon_{ijk}\mbsi{H}_{\alpha j,k}.
\end{align}
\end{subequations}

\begin{remark}[{\bf Notation}]\it
Let ${\nabla_x}, {\mathrm{Div}_x}, {\mathrm{Curl}_x}$ denote the gradient, divergence, and rotation with respect to $\mbs{x}$ keeping $\mbs{y}$, $\mbs{\theta}$ fixed. Similarly, we suppose that
${\nabla_y}, {\mathrm{Div}_y}, {\mathrm{Curl}_y}$ are the gradient, divergence, and rotation with respect to $\mbs{y}$ keeping $\mbs{x}$, $\mbs{\theta}$ fixed and 
${\nabla_\theta}, {\mathrm{Div}_\theta}, {\mathrm{Curl}_\theta}$ are the gradient, divergence, and rotation with respect to $\mbs{\theta}$ keeping $\mbs{x}$, $\mbs{y}$ fixed. 
In the case of indicial notation the same, we adopt the notation such that
the comma followed by a subscript $A$ (resp. $i$ and $\alpha$) denotes the derivative with respect to $x_A$ (resp. $y_i$ and $\theta_\alpha$).
Thus the useful definitions of curl 
are given by
\begin{equation}\begin{aligned}
({{\mathrm{Curl}_x}}\mbf{C})_{iA} = \upepsilon_{A{B}{C}}\mbsi{C}_{iB,C}, \qquad
({{\mathrm{Curl}_y}}\mbf{D})_{Ai} = \upepsilon_{ijk}\mbsi{D}_{Aj,k},\\
({{\mathrm{Curl}_\theta}}\mbf{E})_{A\alpha} = \upepsilon_{\alpha\beta\gamma}E_{A\beta,\gamma}.
\label{defRts}
\end{aligned}\end{equation}
\label{notrem}
\end{remark}

Based on the arguments provide so far, which fulfil the main ingredients of its proof following Olver and Sivaloganathan \cite{OS88}, we state the characterization theorem for null Lagrangians.
\begin{theorem}
The Lagrangian $\sfn{L}$ for the functional of the form \eqref{microlag} is a null Lagrangian if and only if
there exist $\sfn{A},\sfn{D},\wt{\sfn{D}}$ as scalar functions of $(\mbs{x},\mbs{\chi},\mbs{\theta})$,
$\mbf{B},\wt{\mbf{B}},\wh{\mbf{B}}, \mbf{C}, \wt{\mbf{C}}, \wh{\mbf{C}}$ as $3\times3$ matrix functions of $(\mbs{x},\mbs{\chi},\mbs{\theta})$,
and $\mathtt{J}$ as a $3\times3\times3$ matrix function of $(\mbs{x},\mbs{\chi},\mbs{\theta})$ such that
\begin{equation}\begin{aligned}
\sfn{L} &=\sfn{A}+\mbf{B}^{\top} \cdot {\mbf{F}}+\mbf{C}\cdot{{\mathrm{cof~}}} {\mbf{F}}+\sfn{D}\det {\mbf{F}}\\
&+\wt{\mbf{B}}^{\top}\cdot {\mbf{G}}+\wt{\mbf{C}}\cdot{{\mathrm{cof~}}} {\mbf{G}}+\wt{\sfn{D}}\det {\mbf{G}}\\
&+\wh{\mbf{B}}\cdot ({{\mathrm{cof~}}}\mbf{G})\mbf{F}^{\top}+\wh{\mbf{C}}\cdot({{\mathrm{cof~}}}{\mbf{F}}){\mbf{G}}^{\top}+\mathtt{J}\cdot\mbs{e}_\alpha\otimes \mbs{e}_j \otimes(\mbf{G}^{\top}\mbs{e}_\alpha{\wedge}\mbf{F}^{\top}\mbs{e}_j),
\label{E:lt}
\end{aligned}\end{equation}
where the 84 scalar functions appearing as coefficients (or its components) depend only on $36$ scalar functions (as components of $\mbs{L}, \mbs{M}, \wt{\mbs{M}}, \mbf{K}, \wt{\mbf{K}}, \mbf{H}$), in terms of the notation described in Remark \ref{notrem}, in the following way:
\begin{subequations}
\begin{align}
\sfn{A}&={{\mathrm{Div}_x}}\mbs{L}, \qquad\sfn{D}={{\mathrm{Div}_y}}\mbs{M}, \qquad\wt{\sfn{D}}={{\mathrm{Div}_\theta}}\wt{\mbs{M}},\\
\mathtt{J}&=\mbs{e}_\alpha\otimes \mbf{K}_{,\alpha}-\mbs{e}_\alpha\otimes({\nabla_y}\wt{\mbf{K}}^{\top}\mbs{e}_\alpha)^{\top}+{\nabla_x} \mbf{H}\\
\mbf{B}^{\top}&={{\mathrm{Curl}_x}}\mbf{K}+({{\nabla_y}}\mbs{L})^{\top},\qquad \mbf{C}={{\nabla_x}}\mbs{M}- ({{\mathrm{Curl}_y}}\mbf{K}^{\top})^{\top},\\
\wt{\mbf{B}}^{\top}&={{\mathrm{Curl}_x}}\wt{\mbf{K}}+({{\nabla_\theta}}\mbs{L})^{\top},\qquad \wt{\mbf{C}}={{\nabla_x}}\wt{\mbs{M}}- ({{\mathrm{Curl}_\theta}}\wt{\mbf{K}}^{\top})^{\top},\\
\wh{\mbf{B}}&={{\nabla_y}}\wt{\mbs{M}}-({{\mathrm{Curl}_\theta}}{\mbf{H}}^{\top})^{\top},\qquad \wh{\mbf{C}}={{\nabla_\theta}}{\mbs{M}}+({{\mathrm{Curl}_y}}{\mbf{H}})^{\top}.
\end{align}
\label{LMetcfns}
\end{subequations}
Here $\mbf{F}=\nabla{\mbs{\chi}}, \mbf{G}=\nabla{\boldsymbol{\theta}}$, i.e.,
$F_{iA}=\chi_{i,A}, G_{\alpha A}=\theta_{\alpha,A}.$ 
\label{Thm1}
\end{theorem}

\begin{remark}\it
In \eqref{LMetcfns}, $\mbs{L},\mbs{M},\wt{\mbs{M}}$ are 3 component vector functions of $\mbs{x},\mbs{\chi},\mbs{\theta}$;
$\mbf{K},\wt{\mbf{K}}, \mbf{H}$ are $3\times3$ matrix functions of $\mbs{x},\mbs{\chi},\mbs{\theta}$.
In indicial notation, \eqref{E:lt} is alternately expressed as
\begin{equation}\begin{aligned}
\sfn{L}=&\sfn{A}+\mbsi{B}_{Ai}F_{iA}+\mbsi{C}_{iA} ({{\mathrm{cof~}}}\mbf{F})_{iA}+\sfn{D}(\det \mbf{F})\\
&+\wt{\mbsi{B}}_{A\alpha}G_{\alpha A}+\wt{\mbsi{C}}_{iA} ({{\mathrm{cof~}}}\mbf{G})_{iA}+\wt{\sfn{D}}(\det \mbf{G})\\
&+\wh{\mbsi{B}}_{\alpha i}({{\mathrm{cof~}}}\mbf{G})_{\alpha A}F_{iA}+\wh{\mbsi{C}}_{i\alpha}({{\mathrm{cof~}}}\mbf{F})_{i A}G_{\alpha A}
+J_{\alpha j C}G_{\alpha A}F_{jB}\upepsilon_{CAB},
\label{eq3}
\end{aligned}\end{equation}
while, \eqref{LMetcfns} is equivalent to the conditions
\begin{subequations}
\begin{align}
\sfn{A}&=\mbsi{L}_{A,A},\qquad \sfn{D}=\mbsi{M}_{i,i},\qquad \wt{\sfn{D}}=\wt{\mbsi{M}}_{\alpha,\alpha}, \\
J_{\alpha j C}&=(\mbsi{K}_{jC,\alpha}-\wt{\mbsi{K}}_{\alpha C,j}+\mbsi{H}_{\alpha j,C}),\\
\mbsi{B}_{Ai}&
=\mbsi{L}_{A,i}-\upepsilon_{ABC}\mbsi{K}_{iC,B},\qquad
\mbsi{C}_{iA}
=\upepsilon_{ijk} \mbsi{K}_{kA,j}+\mbsi{M}_{i,A},\\
\wt{\mbsi{B}}_{A\alpha}&
=\mbsi{L}_{A,\alpha}-\upepsilon_{ABC}\wt{\mbsi{K}}_{\alpha C,B},\qquad
\wt{\mbsi{C}}_{iA}
=\upepsilon_{i\beta\gamma}\wt{\mbsi{K}}_{\gamma A,\beta}+\wt{\mbsi{M}}_{i,A},\\
\wh{\mbsi{B}}_{\alpha i}
&=\wt{\mbsi{M}}_{A,i}+\upepsilon_{A\beta \gamma}\mbsi{H}_{\gamma i,\beta},\qquad
\wh{\mbsi{C}}_{i\alpha}=\mbsi{M}_{i,\alpha}+\upepsilon_{ijk}\mbsi{H}_{\alpha j,k}.
\end{align}
\label{E:lth}
\end{subequations}
\end{remark}

\begin{remark}\it
Using the characterization of the null Lagrangians 
via the $2$-form \eqref{omegafn}, it is natural to define the class of polyconvex functions relevant for a Cosserat elastic media as follows:
A {\em polyconvex} Lagrangian function for a Cosserat elastic media is given by
\begin{equation}\begin{aligned}
&\Psi\bigg({\mbf{F}},{{\mathrm{cof~}}} {\mbf{F}},\det {\mbf{F}}, {\mbf{G}}, {{\mathrm{cof~}}} {\mbf{G}}, \det {\mbf{G}}, \\
&({{\mathrm{cof~}}}\mbf{G})\mbf{F}^{\top}, ({{\mathrm{cof~}}}{\mbf{F}}){\mbf{G}}^{\top}, \mbs{e}_\alpha\otimes \mbs{e}_j \otimes(\mbf{G}^{\top}\mbs{e}_\alpha{\wedge}\mbf{F}^{\top}\mbs{e}_j)\bigg),
\end{aligned}\end{equation}
where
$\Psi:\mathbb{R}^{83}\to\mathbb{R}$ is a convex function in each of its argument \cite{Ball77,Dac2010}.
Thus, a strain energy density function \eqref{WEK} for a Cosserat elastic media is polyconvex if and only if there exists $\Psi$ of above form.
At this point, it is also useful to list a special sub-class of above via additive decomposition, i.e.,
\begin{equation}\begin{aligned}
&\Phi_1({\mbf{F}})+\Phi_2({{\mathrm{cof~}}} {\mbf{F}})+\Phi_3(\det {\mbf{F}})+\Phi_4({\mbf{G}})+\Phi_5({{\mathrm{cof~}}} {\mbf{G}})+\Phi_6(\det {\mbf{G}})\\
&+\Phi_7(({{\mathrm{cof~}}}\mbf{G})\mbf{F}^{\top})+\Phi_8(({{\mathrm{cof~}}}{\mbf{F}}){\mbf{G}}^{\top})+\Phi_9(\mbs{e}_\alpha\otimes \mbs{e}_j \otimes(\mbf{G}^{\top}\mbs{e}_\alpha{\wedge}\mbf{F}^{\top}\mbs{e}_j)),
\end{aligned}\end{equation}
where the nine functions $\{\Phi_i\}_{i=1}^9$ are convex functions of their arguments.
\end{remark}

\subsection{Divergence representation}
The characterization theorem, stated above as Theorem \ref{Thm1}, can be further applied to obtain a divergence representation of the null Lagrangians akin to Theorem 7 of Olver and Sivaloganathan in \cite{OS88} (see also \cite{Olver86}).
In fact, we find that
\begin{equation}\begin{aligned}
\mbs{P}
&= \mbs{L}+({\mbf{F}}^{\top}\mbf{K})^{\times}+({{\mathrm{cof~}}}{\mbf{F}})^{\top}\mbs{M}+({\mbf{G}}^{\top}\wt{\mbf{K}})^{\times}\\
&+({{\mathrm{cof~}}}{\mbf{G}})^{\top}\wt{\mbs{M}}-\frac{1}{2}(\mbf{F}^{\top}\mbf{H}^{\top}\mbf{G} -\mbf{G}^{\top}\mbf{H}\mbf{F})^{\times},
\label{P3d}
\end{aligned}\end{equation}
where we used the symbolic notation ${}^{\times}$ \eqref{crossM}.
In indicial notation, \eqref{P3d} can be expressed as
\begin{equation}\begin{aligned}
{\mbsi{P}}_A
&= \mbsi{L}_A+\upepsilon_{ABC}{F}_{iB}\mbsi{K}_{iC}+({{\mathrm{cof~}}}{\mbf{F}})_{iA}{M}_i+\upepsilon_{ABC}{G}_{\alpha B}\wt{\mbsi{K}}_{\alpha C}\\
&+({{\mathrm{cof~}}}{\mbf{G}})_{\alpha A}\wt{\mbsi{M}}_\alpha+\upepsilon_{ABC}G_{\alpha B}F_{iC}\mbsi{H}_{\alpha i}.
\end{aligned}\end{equation}
By a direct calculation, it is easy to verify that 
\[
\sfn{L}={\nabla\cdot}\mbs{P}
\]
in \eqref{E:lt} of Theorem above. The detailed steps justifying this identity are provided in Appendix \ref{appC}.

\begin{remark}\it
It is easy to recognize that as a special case of nonlinear elasticity, i.e., absence of effect of microrotation, the expression of $\mbs{P}$ stated in \eqref{P3d} reduces to the well known one (Eq. (13.6.3) of \cite{Silhavy97}), i.e., $\mbs{L}+({\mbf{F}}^{\top}\mbf{K})^{\times}+({{\mathrm{cof~}}}{\mbf{F}})^{\top}\mbs{M}$.
\end{remark}

\section{Null Lagrangian in micropolar shell theory}
In this section, due to a natural presence of curvilinear coordinates (local coordinate chart for two dimensional manifold embedded in three dimensional space \cite{AM78,Ciarlet00,AEReview}), we employ upper and lower indices in this section for contravariant and covariant components \cite{SV06,Kovalev2012}.
\begin{figure}\centering\includegraphics[width=.7\linewidth]{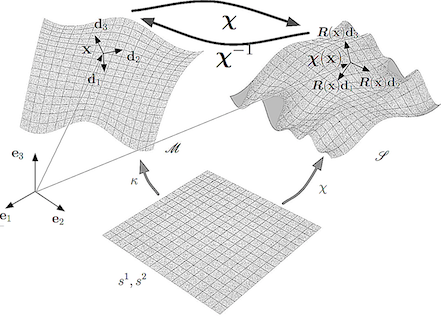}
\caption{Kinematics for a micropolar shell.}
\label{micropolarsurfaces}\end{figure}
It is natural to utilize the parameter space for the shell in place of ${{\si{\ohm}}}$ in this section; this also enables us to avoid the covariant derivative (but it can be easily incorporated by multiplying the Lagrangian by a factor \cite{Kovalev2012}).
It is emphasized that in this section the captial Latin indices $A, B, \dotsc,$ range over $1, 2$.
Here
$(x^1,x^2)$ (in place of the symbols $(s^1, s^2)$ as shown in Fig. \ref{micropolarsurfaces}) are local coordinates on the reference configuration of the shell.
In the assumed framework for micropolar shells, we have the following identification of the local coordinates with components
\[
\mbs{x}=x^A\mbs{e}_A=s^A\mbs{e}_A,\qquad
\mbs{y}=y^i\mbs{e}_i,\qquad
{\boldsymbol{\theta}}=\theta^\alpha\mbs{e}_\alpha.
\]

With $\sfn{A}, \mbf{B}, \mbs{C}, \wt{\mbf{B}}, \wt{\mbs{C}}$, and $\wh{\mbf{B}}$ as local functions of ${x}^B, {y}^h, {\theta}^{\beta},$ 
(so that the total number of the scalar functions is 
$\binom{2+3+3}{2}$, i.e., 
$28$), the general expression of a $2$-form on the shell (as the counterpart of \eqref{omegafn2form}) is found to be
\begin{equation}\begin{aligned}
\label{omegashell}
\omega
&= \sfn{A} dx^1 {{\wedge}} dx^2+\mbsi{B}_{Ai}dy^i {{\wedge}} dx^A+\wt{\mbsi{B}}_{A\alpha}d\theta^\alpha {{\wedge}} dx^A\\
&+\frac{1}{2}\upepsilon_{ijk}\mbsi{C}^i dy^j {{\wedge}} dy^k+\frac{1}{2}\upepsilon_{\alpha\beta\gamma}\wt{\mbsi{C}}^\alpha d\theta^\beta{{\wedge}} d\theta^\gamma+\wh{\mbsi{B}}_{\alpha i}dy^i {{\wedge}} d\theta^\alpha.
\end{aligned}\end{equation}
which leads to a `horizontal' form 
\begin{equation}\begin{aligned}
&
(\sfn{A} +\mbsi{B}_{Ai}y^i_B\epsilon^{AB}+\frac{1}{2}\upepsilon_{ijk}\mbsi{C}^i y^j_A y^k_B\epsilon^{AB}
+\frac{1}{2}\upepsilon_{\alpha\beta\gamma}\wt{\mbsi{C}}^\alpha\theta^\beta_A \theta^\gamma_B \epsilon^{AB}\\
&+\wt{\mbsi{B}}_{A\alpha}\theta^\alpha_B \epsilon^{AB}
+\wh{\mbsi{B}}_{i\alpha}y^i_A \theta^\alpha_B \epsilon^{AB})dx^1 {{\wedge}} dx^2,
\end{aligned}\end{equation}
where
the two dimensional Levi-Civita symbol is denoted by $\upepsilon^{AB}$, i.e.,
\begin{equation}
\upepsilon^{AB}=\begin{cases}
+1&\text{ if } (A,B)=(1,2),\\
-1&\text{ if } (A,B)=(2,1),\\
0&\text{ if } A=B.
\end{cases}
\end{equation}
Similar to the previous section, we continue to use the notation
\begin{equation}\begin{aligned}
\mbf{F}=y^{i}_{,A}\mbs{e}_i\otimes\mbs{e}^A=y^i_A\mbs{e}_i\otimes\mbs{e}^A,\\
\mbf{G}=\theta^\alpha_{,A}\mbs{e}_\alpha\otimes\mbs{e}^A=\theta^\alpha_{A}\mbs{e}_\alpha\otimes\mbs{e}^A,
\end{aligned}\end{equation}
Note that
\begin{equation}
y^i_A\mbs{e}_i=F_{iA}\mbs{e}_i=\mbf{F}\mbs{e}_A={F}_{iB}(\mbs{e}_i\otimes\mbs{e}^B)\mbs{e}_A,
\end{equation}
and similar relation for $\theta^\alpha_A$.
The discussion so far enables us to write the general expression capturing the form of null Lagrangians, as the counterpart of \eqref{omegafn} for the micropolar shell,
\begin{equation}\begin{aligned}
\sfn{L}=\sfn{A}
+\epsilon[\mbf{B}{\mbf{F}}]
+\epsilon[\wt{\mbf{B}}{\mbf{G}}]
+\frac{1}{2}\epsilon^{AB}\mbs{C}\cdot\mbf{F}\mbs{e}_A{\wedge}\mbf{F}\mbs{e}_B
+\frac{1}{2}\epsilon^{AB}\wt{\mbs{C}}\cdot\mbf{G}\mbs{e}_A{\wedge}\mbf{G}\mbs{e}_B\\
+\epsilon^{AB}\wh{\mbf{B}}\cdot(\mbf{F}\mbs{e}_A\otimes\mbf{G}\mbs{e}_B),
\end{aligned}\end{equation}
which can be further simplified to
\begin{subequations}
\begin{equation}\begin{aligned}
\sfn{L} &=\sfn{A}
+\epsilon[\mbf{B}{\mbf{F}}]
+\epsilon[\wt{\mbf{B}}{\mbf{G}}]\\
&+\mbs{C}\cdot\mbf{F}\mbs{e}_1{\wedge}\mbf{F}\mbs{e}_2
+\wt{\mbs{C}}\cdot\mbf{G}\mbs{e}_1{\wedge}\mbf{G}\mbs{e}_2\\
&+\wh{\mbf{B}}\cdot(\mbf{F}\mbs{e}_1\otimes\mbf{G}\mbs{e}_2-\mbf{F}\mbs{e}_2\otimes\mbf{G}\mbs{e}_1),
\end{aligned}\end{equation}
where 
\begin{equation}\begin{aligned}
\mbf{B}&=\mbsi{B}_{Ai}\mbs{e}^A\otimes\mbs{e}^i,
\wt{\mbf{B}}=\wt{\mbsi{B}}_{A\alpha}\mbs{e}^A\otimes\mbs{e}^\alpha,
\wh{\mbf{B}}=\wt{\mbsi{B}}_{i\alpha}\mbs{e}^i\otimes\mbs{e}^\alpha,\\
\mbs{C}&=\mbsi{C}^{i}\mbs{e}_i,
\wt{\mbs{C}}=\wt{\mbsi{C}}^{\alpha}\mbs{e}_\alpha,
\end{aligned}\end{equation}
and
\begin{equation}\begin{aligned}
\epsilon[\mbf{A}]=\mbf{A}\mbs{e}_2\cdot\mbs{e}_1-\mbf{A}\mbs{e}_1\cdot\mbs{e}_2.
\end{aligned}\end{equation}
\label{omegafshell}
\end{subequations}
The coordinate expression of the relevant $1$-form $\zeta$ (as counterpart of \eqref{zeta3d}) is written as
\begin{equation}
\zeta = \ol{\mbsi{P}}_Ad{x}^A + \wh{\mbsi{P}}_kd{y}^k + \wt{\mbsi{P}}_{\alpha}d{\theta}^\alpha,
\label{zetadefshell}
\end{equation}
where $\ol{\mbsi{P}}_A, \wh{\mbsi{P}}_k$, and $\wt{\mbsi{P}}_{\alpha}$ are local functions of ${x}^B, {y}^h, {\theta}^{\beta}.$ 
The total number of the scalar functions is $
8$, as expected $\binom{2+3+3}{2-1}
$.
Using this expression, the exterior derivative of $\zeta$ \eqref{zetadefshell} can be written as
\begin{equation}
\begin{aligned}
d\zeta=\pd{\ol{\mbsi{P}}_A}{{x}^B} d{x}^B{\wedge}d{x}^A+\pd{\wt{\mbsi{P}}_{\alpha}}{{\theta}^\beta}d{\theta}^{\beta}{\wedge}d{\theta}^\alpha+\pd{\wh{\mbsi{P}}_k}{{y}^h}d{y}^h{\wedge}d{y}^k+ (\pd{\ol{\mbsi{P}}_A}{{\theta}^\alpha} -\pd{\wt{\mbsi{P}}_{\alpha}}{{x}^A})d{\theta}^\alpha{\wedge}d{x}^A\\
+ (\pd{\ol{\mbsi{P}}_A}{{y}^k} -\pd{\wh{\mbsi{P}}_{k}}{{x}^A})d{y}^k{\wedge}d{x}^A+ (\pd{\wt{\mbsi{P}}_\alpha}{{y}^k} -\pd{\wh{\mbsi{P}}_{k}}{{\theta}^\alpha})d{y}^k{\wedge}d{\theta}^\alpha.
\end{aligned}
\label{dzetashell1}
\end{equation}
Using above expression of $d\zeta$ and the expession of $\omega$ 
\eqref{omegashell},
we find that the conditions corresponding to a null Lagrangian are
\begin{equation}
\begin{aligned}
\sfn{A}=\pd{\ol{\mbsi{P}}_A}{{x}^B}\upepsilon_{BA},
\mbsi{B}_{Ai}=(\pd{\ol{\mbsi{P}}_A}{{y}^i} -\pd{\wh{\mbsi{P}}_{i}}{{x}^A}),
\wt{\mbsi{B}}_{A\alpha}=(\pd{\ol{\mbsi{P}}_A}{{\theta}^\alpha} -\pd{\wt{\mbsi{P}}_{\alpha}}{{x}^A}),\\
\frac{1}{2}\upepsilon_{ijk}\mbsi{C}^i =\pd{\wh{\mbsi{P}}_k}{{y}^j},
\frac{1}{2}\upepsilon_{\alpha\beta\gamma}\wt{\mbsi{C}}^\alpha=\pd{\wt{\mbsi{P}}_{\gamma}}{{\theta}^\beta},
\wh{\mbsi{B}}_{\alpha i}=(\pd{\wt{\mbsi{P}}_\alpha}{{y}^i} -\pd{\wh{\mbsi{P}}_{i}}{{\theta}^\alpha}).
\end{aligned}
\label{nullshellcond1}
\end{equation}
In direct notation, the set of conditions \eqref{nullshellcond1} can be re-written as
\begin{equation}\begin{aligned}
\sfn{A}=\ol{\mbsi{P}}_{1,2}-\ol{\mbsi{P}}_{2,1}=\epsilon[{\nabla_x}\mbs{\ol{P}}],\\
\mbf{B}={\nabla_y}\mbs{\ol{P}}-({\nabla_x}\mbf{\wh{P}})^{\top}, \wt{\mbf{B}}={\nabla_\theta}\mbs{\ol{P}}-({\nabla_x}\mbs{\wt{P}})^{\top},\\
{\mbs{C}}={\mathrm{Curl}_y}\mbs{\wh{P}}, \wt{\mbs{C}}={\mathrm{Curl}_\theta}\mbs{\wt{P}},\\
\wh{\mbf{B}}=\pd{\wt{\mbsi{P}}_\alpha}{{y}^i}\mbs{e}^\alpha\otimes\mbs{e}^i -\pd{\wh{\mbsi{P}}_{i}}{{\theta}^\alpha}\mbs{e}^\alpha\otimes\mbs{e}^i
={\nabla_y}\mbs{\wt{P}}-({\nabla_\theta}{\mbs{\wh{P}}})^{\top}.
\end{aligned}\end{equation}

\begin{theorem}
The Lagrangian $\sfn{L}$ for the functional of the form \eqref{microlag} for a micropolar shell is a null Lagrangian if and only if
there exist $\sfn{A}$ as a scalar functions of $(\mbs{x},\mbs{\chi},\mbs{\theta})$,
$\mbf{B},  \wt{\mbf{B}}$ as $2\times3$ matrix functions of $(\mbs{x},\mbs{\chi},\mbs{\theta})$,
$\wh{\mbf{B}}$ as $3\times3$ matrix function of $(\mbs{x},\mbs{\chi},\mbs{\theta})$,
and $\mbs{C}$ and $\wt{\mbs{C}}$ as a $3$ component vector functions of $(\mbs{x},\mbs{\chi},\mbs{\theta})$ such that
\eqref{omegafshell} holds
where the 28 scalar functions appearing as coefficients (or its components) depend only on $8$ scalar functions (as components of $\mbs{\ol{P}}, \mbs{\wh{P}}, \mbs{\wt{P}}$)
in the following way:
\begin{subequations}
\begin{align}
\sfn{A}&=\epsilon[{\nabla_x}\mbs{\ol{P}}],\qquad
\mbf{B}={\nabla_y}\mbs{\ol{P}}-({\nabla_x}\mbf{\wh{P}})^{\top},\\
\wt{\mbf{B}}&={\nabla_\theta}\mbs{\ol{P}}-({\nabla_x}\mbs{\wt{P}})^{\top},\qquad
\wh{\mbf{B}}={\nabla_y}\mbs{\wt{P}}-({\nabla_\theta}{\mbs{\wh{P}}})^{\top},\\
{\mbs{C}}&={\mathrm{Curl}_y}\mbs{\wh{P}},\qquad
\wt{\mbs{C}}={\mathrm{Curl}_\theta}\mbs{\wt{P}}.
\end{align}
\label{LMetcfnshell}
\end{subequations}
\label{Thm1shell}
\end{theorem}

\begin{remark}\it
Using the characterization of the null Lagrangians \eqref{omegafshell}
via the $2$-form \eqref{omegashell}, it is natural to define the class of polyconvex functions for micropolar shells \cite{Ciarlet2011,Birsan2014} as follows.
A polyconvex lagrangian for a micropolar shell is given by
\begin{equation}\begin{aligned}
\sfn{L}(\mbs{x},{\mbs{\chi}},{\boldsymbol{\theta}},\nabla{\mbs{\chi}},\nabla{\boldsymbol{\theta}})&:=\Psi\bigg({\mbf{F}}, {\mbf{G}},
\mbf{F}\mbs{e}_1{\wedge}\mbf{F}\mbs{e}_2,
\mbf{G}\mbs{e}_1{\wedge}\mbf{F}\mbs{e}_2,\\
&\mbf{F}\mbs{e}_1\otimes\mbf{G}\mbs{e}_2-\mbf{F}\mbs{e}_2\otimes\mbf{G}\mbs{e}_1\bigg),
\end{aligned}\end{equation}
where
$\Psi:\mathbb{R}^{
27}\to\mathbb{R}$ is a convex function in each of its argument \cite{Ball77,Dac2010}.
\end{remark}

\subsection{Divergence representation}
We expect to reduce above expression of $\sfn{L}$ in Theorem \ref{Thm1shell}
as $P_{1,1}+P_{2,2}$ for a vector $\mbs{P}\sim(P_1,P_2)$.
The expression 
\eqref{dzetashell1} leads to a `horizontal' form 
\begin{equation}
\begin{aligned}
\bigg(\pd{\ol{\mbsi{P}}_B}{{x}^A}+(\pd{\ol{\mbsi{P}}_B}{{\theta}^\alpha} -\pd{\wt{\mbsi{P}}_{\alpha}}{{x}^B}){\theta}^\alpha_{,A}+(\pd{\ol{\mbsi{P}}_B}{{y}^k} -\pd{\wh{\mbsi{P}}_{k}}{{x}^B}){y}^k_{,A}\\
+\pd{\wt{\mbsi{P}}_{\alpha}}{{\theta}^\beta}{\theta}^\beta_{,A}{\theta}^\alpha_{,B}+(\pd{\wt{\mbsi{P}}_\alpha}{{y}^k} -\pd{\wh{\mbsi{P}}_{k}}{{\theta}^\alpha}){y}^k_{,A}{\theta}^\alpha_{,B}+\pd{\wh{\mbsi{P}}_k}{{y}^h}{y}^h_{,A} {y}^k_{,B}\bigg)d{x}^A\wedge d{x}^B.
\end{aligned}
\label{horzzetashell}
\end{equation}
Using \eqref{horzzetashell}, the local expression of a null Lagrangian for micropolar shells is found to be
\begin{equation}
\begin{aligned}
\sfn{L}=\bigg(\pd{\ol{\mbsi{P}}_B}{{x}^A}+(\pd{\ol{\mbsi{P}}_B}{{\theta}^\alpha} -\pd{\wt{\mbsi{P}}_{\alpha}}{{x}^B}){\theta}^\alpha_{,A}+(\pd{\ol{\mbsi{P}}_B}{{y}^k} -\pd{\wh{\mbsi{P}}_{k}}{{x}^B}){y}^k_{,A}\\
+\pd{\wt{\mbsi{P}}_{\alpha}}{{\theta}^\beta}{\theta}^\beta_{,A}{\theta}^\alpha_{,B}+(\pd{\wt{\mbsi{P}}_\alpha}{{y}^k} -\pd{\wh{\mbsi{P}}_{k}}{{\theta}^\alpha}){y}^k_{,A}{\theta}^\alpha_{,B}+\pd{\wh{\mbsi{P}}_k}{{y}^h}{y}^h_{,A} {y}^k_{,B}\bigg)\epsilon^{AB}.
\end{aligned}
\end{equation}
Indeed, (with $|$ as a decoration to denote the `total' derivative) by a repeated application of the product rule and chain rule of differentiation,
\begin{equation}
\begin{aligned}
\sfn{L}&=\bigg(\pd{}{{x}^A}|\ol{\mbsi{P}}_B
-\pd{\wt{\mbsi{P}}_{\alpha}}{{x}^B}{\theta}^\alpha_{,A}-\pd{\wh{\mbsi{P}}_{k}}{{x}^B}{y}^k_{,A}\\
&-\pd{\wt{\mbsi{P}}_{\alpha}}{{x}^A}{\theta}^\alpha_{,B}+\pd{}{{x}^A}|\wt{\mbsi{P}}_{\alpha}{\theta}^\alpha_{,B}+(-\pd{\wh{\mbsi{P}}_{k}}{{\theta}^\alpha}){y}^k_{,A}{\theta}^\alpha_{,B}-\pd{\wh{\mbsi{P}}_k}{{y}^h}{y}^h_{,B} {y}^k_{,A}\bigg)\epsilon^{AB},
\end{aligned}
\end{equation}
which can be further written as
\begin{equation}
\begin{aligned}
\sfn{L}&=\bigg(\pd{}{{x}^A}|\ol{\mbsi{P}}_B+\pd{}{{x}^A}|\wt{\mbsi{P}}_{\alpha}{\theta}^\alpha_{,B}+\pd{}{{x}^A}|\wh{\mbsi{P}}_{k}{y}^k_{,B}\bigg)\epsilon^{AB}\\
&=\pd{}{{x}^A}|\bigg(\ol{\mbsi{P}}_B+\wt{\mbsi{P}}_{\alpha}{\theta}^\alpha_{,B}+\wh{\mbsi{P}}_{k}{y}^k_{,B}\bigg)\epsilon^{AB}.
\end{aligned}
\label{nulllagdiv1}
\end{equation}
Above expression \eqref{nulllagdiv1} motivates the definition
\begin{equation}
\begin{aligned}
P_A:=\epsilon^{AB}(\ol{\mbsi{P}}_B+\wt{\mbsi{P}}_{\alpha}{\theta}^\alpha_{,B}+\wh{\mbsi{P}}_{k}{y}^k_{,B}).
\end{aligned}
\end{equation}
Thus, in direct notation,
\begin{equation}
\begin{aligned}
\sfn{L}={\nabla\cdot}\mbs{P},\quad \mbs{P}=\epsilon[\mbs{\ol{P}}+\mbf{F}^{\top}\mbs{\wh{P}}+\mbf{G}^{\top}\mbs{\wt{P}}],
\end{aligned}
\end{equation}
where $\mbs{P}=\mbs{P}(\mbs{x}, \mbs{y}(\mbs{x}), \mbs{\theta(\mbs{x})})$, and $\mbs{\ol{P}}, \mbs{\wh{P}}, \mbs{\wt{P}}$ are functions of $\mbs{x}, \mbs{y}, \mbs{\theta}$.

\section{Nilpotent energies in linearized micropolar theory}
In the special case of 
homogenous, linearized micropolar theory \cite{Er66,Eringen99,Alt2012}, 
using the standard notation for fourth order tensors \cite{Gurtin1981}, we are looking for Lagrangians of the form
\begin{equation}\begin{aligned}
&\frac{1}{2}({\nabla}\mbs{u}+\epsilon\mbs{\phi})\cdot\mathbb{A}({\nabla}\mbs{u}+\epsilon\mbs{\phi})+\frac{1}{2}{\nabla}\mbs{\phi}\cdot\mathbb{B}{\nabla}\mbs{\phi}\\
&+({\nabla}\mbs{u}+\epsilon\mbs{\phi})\cdot\mathbb{D}{\nabla}\mbs{\phi},
\label{E:lo}
\end{aligned}\end{equation}
with $\mathbb{A},\mathbb{B},\mathbb{D}$ constant tensors such that 
\begin{equation}\begin{aligned}
\label{ABsym}
{A}_{ijkl}={A}_{klij}, \qquad
{B}_{ijkl}={B}_{klij};
\end{aligned}\end{equation}
in indicial notation, 
\[
\frac{1}{2} {A}_{ijkl}(u_{i,j}+{\epsilon}_{ijs}\phi_s)(u_{k,l}+{\epsilon}_{klt}\phi_t)
+\frac{1}{2} {B}_{ijkl} {\phi}_{ij}{\phi}_{kl}
+{D}_{ijkl}(u_{i,j}+{\epsilon}_{ijs}\phi_s){\phi}_{kl}.
\]
Here $\mbs{u}$ represents the (infinitesimal) displacement field while $\mbs{\phi}$ represents the (infinitesimal) rotation vector field.
In the context of the Lagrangian $\sfn{L}$ in \eqref{microlag}, with ${\mbs{\chi}}$ (resp. ${\boldsymbol{\theta}}$) replaced by $\mbs{u}$ (resp. $\mbs{\phi}$), 
according to the characterization theorem for first order null Lagrangians in the micropolar theory, by comparison of \eqref{E:lo} and \eqref{E:lt}, 
thus, we find that only terms bilinear in $\mbf{F}=\nabla\mbs{u}$ and $\mbs{\phi}$ and quadratic in $\mbf{F}, \mbf{G}=\nabla\mbs{\phi}$ (and $\mbs{\phi}$) are needed as well as mixed type where terms are bilinear in $\mbf{F}, \mbf{G}$, and $\mbf{G}, \mbs{\phi}$ as well.
Indeed, we conclude that 
$\sfn{A}(\mbs{x},\mbs{u},\mbs{\phi})$ is bilinear in $\mbs{\phi}$ and independent of $\mbs{u}$ and $\mbs{x}$,
$\sfn{D}(\mbs{x},\mbs{u},\mbs{\phi})=\wt{\sfn{D}}(\mbs{x},\mbs{u},\mbs{\phi})=0,$
$\mbf{B}(\mbs{x},\mbs{u},\mbs{\phi}),\wt{\mbf{B}}(\mbs{x},\mbs{u},\mbs{\phi})$ is linear in $\mbs{\phi}$ and independent of $\mbs{u}$ and $\mbs{x}$,
$\mbf{C}(\mbs{x},\mbs{u},\mbs{\phi}),\wt{\mbf{C}}(\mbs{u},\mbs{\phi})$ is a constant tensor,
$\wh{\mbf{B}}(\mbs{x},\mbs{u},\mbs{\phi})=\wh{\mbf{C}}(\mbs{x},\mbs{u},\mbs{\phi})=0$,
$\mathtt{J}(\mbs{x},\mbs{u},\mbs{\phi})$ is a constant tensor.

Resorting to the indicial notation prescribed in Remark \ref{notrem},
here
$F_{iA}=u_{i,A}, G_{\alpha A}=\phi_{\alpha,A},$ and $\mbs{e}_A, \mbs{e}_j, \mbs{e}_\alpha$ are used to denote the same basis vectors $\mbs{e}_1, \mbs{e}_2, \mbs{e}_3$.
However, in the following sometimes we follow ordinary indicial notation while other times we stick to Remark \ref{notrem}.

\begin{theorem}
A function of the form \eqref{E:lo} (with the conditions \eqref{ABsym}) is a null Lagrangian if and only if
$\mathbb A$, $\mathbb B$ and $\mathbb D$ satisfy
(no sum for last two conditions)
\begin{equation}
\label{E:lcon}
\begin{split}
\mathbb A&=0\\
B_{ijkl}&=-B_{ilkj},\\
D_{ijkl}&=-D_{ilkj},\\
D_{ijji}&=-D_{ikki}\text{ for }i\ne j\ne k\ne i,\\
D_{ijjk}&=D_{kikk}+D_{jijk}\text{ for }i\ne j\ne k\ne i.
\end{split}
\end{equation}
\label{linthm}
\end{theorem}

For the proof of above theorem, the sufficiency can be checked by direct substitution in the Euler--Lagrange equations; the details of the same are also provided in \cite{Basak1}.
Therefore, the only non-trivial part is the necessity which we establish in the following.

Using \eqref{cofAcomp},
$\mbsi{C}_{iA}({\mathrm{cof~}} F)_{iA}=\mbsi{C}_{ij}\frac{1}{2}\upepsilon_{imn}\upepsilon_{jpq}F_{mp}F_{nq}=\mbsi{C}_{ij}\frac{1}{2}\upepsilon_{imn}\upepsilon_{jpq}u_{m,p}u_{n,q},$
i.e.,
\begin{equation}
A_{iAjB}=\mbsi{C}_{mn}\frac{1}{2}\upepsilon_{mij}\upepsilon_{nAB}.
\label{Arel1}
\end{equation}
\begin{equation}\begin{aligned}
\implies A_{iAjB}+A_{iBjA}=0,\forall i,j,A,B,\\
\text{ in particular, }A_{iAjA}=0, A_{iAiB}=0.
\label{Arel2}
\end{aligned}\end{equation}
Similarly, a similar argument leads to the result for $\mathbb{B}$, i.e.,
\begin{equation}\begin{aligned}
\implies B_{iAjB}+B_{iBjA}=0,\\
\text{ in particular, }B_{iAjA}=0, \forall i,j,A,
B_{iAiB}=0,\forall i,A,B.
\end{aligned}\end{equation}
Comparing \eqref{E:lo} with \eqref{E:lt} and using \eqref{E:lth} we get
\begin{align*}
D_{ij\alpha A}e_{ijs}\phi_s\phi_{\alpha, A}=&(\mbsi{L}_{A,\alpha}-e_{ABC}\tilde{K}_{\alpha C,B})\phi_{\alpha, A},\\
D_{ij\alpha A}e_{ijs}\phi_s=&\mbsi{L}_{A,\alpha}-e_{ABC}\tilde{K}_{\alpha C,B}\\
\therefore\qquad \mbsi{L}_{A,\alpha A}=&e_{ABC}\tilde{K}_{\alpha C,BA}=e_{ABC}\tilde{K}_{\alpha C,AB}\\
=&-e_{BAC}\tilde{K}_{\alpha C,AB}=-e_{ABC}\tilde{K}_{\alpha C,BA}=0.
\end{align*}
Now, $\mbsi{L}_{A,A}=\frac{1}{2}A_{ijkl}e_{ijs}e_{klt}\phi_s\phi_t$,
\begin{gather}
\therefore\qquad A_{ijkl}e_{ijs}e_{kl\alpha}\phi_s=\mbsi{L}_{A,A\alpha}=\mbsi{L}_{A,\alpha A}=0\qquad\forall\alpha\label{E:lexx}.
\end{gather}
We fix $\beta\in\{1,2,3\}$, then we put
\begin{equation}
\phi_s=
\begin{cases}
0&\text{if }s\ne\beta\\
1&\text{if }s=\beta
\end{cases}.
\end{equation}
Therefore we get $(A_{ijkl}-A_{jikl})e_{kl\alpha}=0 \forall\alpha (i\ne j\ne\beta\ne i)$. Putting $\alpha=i$, the only possibility for $k,l$ is $\beta,j,$ (no sum)
\begin{gather}
A_{ij\beta j}-A_{ji\beta j}-A_{ij j\beta}+A_{ji j\beta}=0\\
\eqref{Arel2}{}_2,\eqref{Arel2}{}_3
\implies A_{ijj\beta}+A_{ji\beta j}=0\notag\\
\therefore\qquad A_{j\beta ij}=-A_{ji\beta j}.
\label{E:lex}
\end{gather}
Again comparing \eqref{E:lo} with \eqref{E:lt} we get
\begin{gather}
A_{iAkl}e_{kls}\phi_su_{i,A}=(\mbsi{L}_{A,i}-e_{ABC}\mbsi{K}_{iC,B})u_{i,A}\notag\\
\implies A_{iAkl}e_{kls}\phi_s=\mbsi{L}_{A,i}-e_{ABC}\mbsi{K}_{iC,B}\label{E:ltr}\\
D_{ij\alpha A}e_{ijs}\phi_s=\mbsi{L}_{A,\alpha}-e_{ABC}\tilde{K}_{\alpha C,B}\label{E:lft}\\
J_{\alpha jC}G_{\alpha A}F_{jB}e_{CAB}=D_{jB\alpha A}u_{j,B}\phi_{\alpha,A},\notag\\
D_{jE\alpha F}=e_{CFE}(\mbsi{K}_{jC,\alpha}-\tilde{K}_{\alpha C,j}+\mbsi{H}_{\alpha j,C})\label{E:lfft}.
\end{gather}
From \eqref{E:lfft}, we have 
\begin{equation}
D_{ijkl}=-D_{ilkj}.
\end{equation}
From \eqref{E:lfft} we get
\begin{equation}
\label{E:lst}
D_{jA\alpha A}=0\qquad\forall\alpha,j,A.
\end{equation}

Also if $E\ne F$ then $C$ has exactly one value for which RHS of \eqref{E:lfft} is nonzero. So there is no sum on $C$. Now differentiating \eqref{E:lfft} with respect to $x_B$ we get
\begin{equation}
\mbsi{K}_{jC,\alpha B}-\tilde{K}_{\alpha C,jB}+\mbsi{H}_{\alpha j,CB}=0
\end{equation}
But 
\begin{align*}
e_{ABC}\mbsi{H}_{\alpha j,CB}&=e_{ABC}\mbsi{H}_{\alpha j,BC}
=-e_{ACB}\mbsi{H}_{\alpha j,BC}=-e_{ABC}\mbsi{H}_{\alpha j,CB}=0.
\end{align*}
Therefore, 
\begin{equation}
e_{ABC}\mbsi{K}_{iC,\alpha B}=e_{ABC}\tilde{K}_{\alpha C,iB}.
\end{equation}
Now differentiating \eqref{E:ltr} with respect to $\phi_\alpha$ and \eqref{E:lft} with respect to $u_i$ we get \begin{gather*}
\mbsi{L}_{A,i\alpha}-A_{iAkl}e_{kl\alpha}=e_{ABC}\mbsi{K}_{iC,\alpha B}=e_{ABC}\tilde{K}_{\alpha C,iB}=\mbsi{L}_{A,\alpha i}\\
\implies A_{iAkl}e_{kl\alpha}=0\qquad\text{as }\mbsi{L}_{A,\alpha i}=\mbsi{L}_{A,i\alpha}\\
\implies A_{iAkl}=A_{iAlk}\qquad\forall i,A,k,l\\
\eqref{Arel1}\implies A_{ilkA}=A_{iklA}\qquad\forall i,A,k,l\\
\therefore\qquad A_{ilki}=A_{ikli}=-A_{ilki}=0\qquad\text{by \eqref{E:lex}}\\
\implies A_{lijl}=0=A_{jlli}\qquad\forall i,j,l\\
\implies A_{llji}=0=A_{jill}\qquad\forall i,j,l.
\end{gather*}
Therefore we get $A_{ijkl}=0$ if any two of its subscript are equal as (no sum)
\begin{equation}
A_{llij}=0, A_{lilj}=0, A_{lijl}=0, A_{illj}=0, A_{iljl}=0, A_{ijll}=0,
\end{equation}
where in view of the symmetry $A_{llij}=A_{ijll}$ and $A_{illj}=A_{ljil}$.
Hence $\mathbb A=0$.

We expand and rewrite \eqref{E:lo} as
\begin{equation}\begin{aligned}
\sfn{L}&=\frac{1}{2}A_{ijkl}u_{i,j}u_{k,l}+A_{ijkl}e_{ijs}\phi_su_{k,l}+\frac{1}{2}A_{ijkl}e_{ijs}e_{klt}\phi_s\phi_t+\frac{1}{2}B_{ijkl}\phi_{i,j}\phi_{k,l}\\
&+D_{ijkl}u_{i,j}\phi_{k,l}+D_{ijkl}e_{ijs}\phi_s\phi_{k,l}\label{E:ltw}\\
&=: \sfn{L}_1+\sfn{L}_2+\sfn{L}_3+\sfn{L}_4+\sfn{L}_5+\sfn{L}_6,
\end{aligned}\end{equation}
where the terms are assigned sequenctially.
Applying the Euler operator  (the sum on index $t$ ranges over $1,2,3$)
\[
\mathscr E_r:=\frac{\partial}{\partial z_r}-\frac{d}{dx_t}\left(\frac{\partial}{\partial p_{rt}}\right)\qquad
(\text{with }r=1,2,3\text{ for }\mbf u; r=4,5,6\text{ for }\boldsymbol\phi)
\]
on \eqref{E:ltw}, where $z$ refers to components of $\mbf u$ and $\boldsymbol\phi$ while $p$ refers to the components of their gradients. 
Then for $r=1,\dots,6,$ $\mathscr E_r(\sfn{L}_1)=\mathscr E_r(\sfn{L}_2)=\mathscr E_r(\sfn{L}_3)=0$ identically, as $\mathbb A=0$; and $\mathscr E_r(\sfn{L}_4)\equiv0.$
So if \eqref{E:ltw} is a null Lagrangian then for $r=1,\dots,6,$ $\mathscr E_r(\sfn{L}_5+\sfn{L}_6)\equiv0$. For $r=1,2,3$,
\begin{align}
\mathscr E_r(\sfn{L}_5+\sfn{L}_6)=\mathscr E_r(\sfn{L}_5)&\equiv0\notag\\
\implies D_{rtkl}\phi_{k,lt}&\equiv0.
\label{E:po}
\end{align}
Fix $i,j,h\in\{1,2,3\}$. Put $\phi_{h,ij}=1$ (so that
$\phi_{h,ji}=1$), while let other components of $\phi_{k,lt}=0$. Then from \eqref{E:po} we get, $D_{rikj}+D_{rjki}=0$ which implies
\begin{gather}
D_{rikj}=-D_{rjki}\label{E:pt}\\
\implies D_{riki}=0.
\label{E:pex}
\end{gather}
Now for $r=4,5,6,$ (h=r-3),
\begin{align}
\mathscr E_r(\sfn{L}_5+\sfn{L}_6)\equiv0,
\implies D_{ijht}u_{i,jt}+D_{ijht}e_{ijs}\phi_s,t-D_{ijkl}e_{ijh}\phi_{k,l}\equiv0.
\label{E:pth}
\end{align}
With proper choice of $\mbf u$ we again get \eqref{E:pt}.
Therefore, as a result of \eqref{E:pth}, we get
\begin{gather}
\left(D_{ijht}e_{ijs}-D_{ijst}e_{ijh}\right)\phi_{s,t}\equiv0,\\
\text{ i.e.},\qquad D_{ijht}e_{ijs}=D_{ijst}e_{ijh}\qquad\forall h,s,t.
\label{E:pf}
\end{gather}
With $(h,s,t)=\pi(1,2,3)$, with the notation that $\pi$ stands for the circular permutation map, then \eqref{E:pf} gives (no sum)
\begin{gather}
D_{thht}-D_{htht}=D_{stst}-D_{tsst}\\
\implies D_{thht}=-D_{tsst}\qquad\text{by \eqref{E:pex}}.
\label{E:pff}
\end{gather}
Also if $(h,s,t)=-\pi(1,2,3)$ then by same calculation \eqref{E:pff} holds.
With $s=t\ne h$, we assume that $(s,h,k)=\pi(1,2,3)$. Then \eqref{E:pf} gives (no sum)
\begin{gather}
D_{hkhs}-D_{khhs}=D_{ksss}-D_{skss}\\
\implies D_{khhs}=D_{hkhs}+D_{skss}\qquad\text{by \eqref{E:pex}}\label{E:ps}
\end{gather}
Also if $(s,h,k)=-\pi(1,2,3)$ then by same calculation \eqref{E:ps} holds. Also for $h=t\ne s$, by similar argument \eqref{E:ps} holds.

\begin{remark}\it
Some other aspects pertaining to the sufficiency part of the Theorem \ref{linthm} are discussed in \cite{Basak1} along with a
few other generalizations of linearized theory of elasticity.
\end{remark}

\section*{Acknowledgments}
BLS gratefully acknowledges the partial support of SERB MATRICS grant MTR/2017/000013.

\appendix
\section{Expansion of $\omega$}
\label{appA}
For the purpose of convenience of derivation, we assume that $\det\mbf{G}\ne0$, let $\nabla_\theta {\mbs{\chi}}=\mbf{T}=\mbf{F}\mbf{G}^{-1}$, then
\begin{equation}
F_{iA}=y_{i,A},
G_{\alpha A}=\theta_{\alpha,A}.
\end{equation}
Then
\begin{equation}\begin{aligned}
\omega
&= \sfn{A} \Omega_v+\frac{1}{2}\upepsilon_{ABC}\mbsi{B}_{Ai}F_{iM}dx_M {{\wedge}} dx_B {{\wedge}} dx_C+\frac{1}{2}\upepsilon_{ijk}\mbsi{C}_{iA}dy_j{{\wedge}} dy_k {{\wedge}} (\mbf{F}^{-1})_{Al}dy_l+\sfn{D}dy_1 {{\wedge}} dy_2 {{\wedge}} dy_3\\
&+\frac{1}{2}\upepsilon_{ABC}\wt{\mbsi{B}}_{A\alpha}G_{\alpha M}dx_M {{\wedge}} dx_B {{\wedge}} dx_C+\frac{1}{2}\upepsilon_{i\beta\gamma}\wt{\mbsi{C}}_{iA}d\theta_\beta{{\wedge}} d\theta_\gamma {{\wedge}} (\mbf{G}^{-1})_{A\alpha}d\theta_\alpha+\wt{\sfn{D}}d\theta_1 {{\wedge}} d\theta_2 {{\wedge}} d\theta_3\\
&+\frac{1}{2}\upepsilon_{\alpha\beta\gamma}\wh{\mbsi{B}}_{\alpha i}T_{i\alpha}d\theta_\alpha {{\wedge}} d\theta_\beta {{\wedge}} d\theta_\gamma+\frac{1}{2}\upepsilon_{ijk}\wh{\mbsi{C}}_{i\alpha}dy_j{{\wedge}} dy_k {{\wedge}} (\mbf{T}^{-1})_{\alpha l}dy_l\\
&+J_{\alpha j C}G_{\alpha A}F_{jB}dx_A {{\wedge}} dx_B {{\wedge}} dx_C,
\end{aligned}\end{equation}
where
\begin{equation}
\label{omgv}
\Omega_v:=dx_1 {{\wedge}} dx_2 {{\wedge}} dx_3.
\end{equation}
Expanding further,
\begin{equation}\begin{aligned}
\omega
&= \sfn{A} \Omega_v
+\frac{1}{2}\upepsilon_{ABC}\upepsilon_{MBC}\mbsi{B}_{Ai}F_{iM}\Omega_v+\frac{1}{2}\upepsilon_{ijk}\mbsi{C}_{iA} (\mbf{F}^{-1})_{Al} \upepsilon_{jkl}(\det \mbf{F})\Omega_v+\sfn{D}(\det \mbf{F})\Omega_v\\
&+\frac{1}{2}\upepsilon_{ABC}\upepsilon_{MBC}\wt{\mbsi{B}}_{A\alpha}G_{\alpha M}\Omega_v+\frac{1}{2}\upepsilon_{i\beta\gamma}\wt{\mbsi{C}}_{iA} (\mbf{G}^{-1})_{A\alpha} \upepsilon_{\alpha\beta\gamma}(\det \mbf{G})\Omega_v+\wt{\sfn{D}}(\det \mbf{G})\Omega_v\\
&+\frac{1}{2}\upepsilon_{\delta\beta\gamma}\upepsilon_{\alpha\beta\gamma}\wh{\mbsi{B}}_{\delta i}T_{i \alpha}(\det\mbf{G})\Omega_v+\frac{1}{2}\upepsilon_{ijk}\wh{\mbsi{C}}_{i\alpha} (\mbf{T}^{-1})_{\alpha l} \upepsilon_{jkl}(\det \mbf{F})\Omega_v\\
&+\upepsilon_{ABC} J_{\alpha j C}G_{\alpha A}F_{jB}\Omega_v.
\end{aligned}\end{equation}
Simplifying above expression, we find that
\begin{equation}\begin{aligned}
\omega
&= (\sfn{A}+\mbsi{B}_{Ai}F_{iA}+\mbsi{C}_{iA} (\mbf{F}^{-1})_{Ai} (\det \mbf{F})+\sfn{D}(\det \mbf{F})\\
&+\wt{\mbsi{B}}_{A\alpha}G_{\alpha A}+\wt{\mbsi{C}}_{\alpha A} (\mbf{G}^{-1})_{A\alpha} (\det \mbf{G})+\wt{\sfn{D}}(\det \mbf{G})\\
&+\wh{\mbsi{B}}_{\alpha i}T_{i\alpha}\det\mbf{G}+\wh{\mbsi{C}}_{i\alpha} (\mbf{T}^{-1})_{\alpha i} (\det \mbf{F})
+J_{\alpha j C}G_{\alpha A}F_{jB}\upepsilon_{CAB})\Omega_v,
\end{aligned}\end{equation}
which can be also written as
\begin{equation}\begin{aligned}\omega&=(\sfn{A}+\mbf{B}^{\top}\cdot {\mbf{F}}+\mbf{C}\cdot{{\mathrm{cof~}}} {\mbf{F}}+\sfn{D}\det {\mbf{F}}+\wt{\mbf{B}}^{\top}\cdot {\mbf{G}}+\wt{\mbf{C}}\cdot{{\mathrm{cof~}}} {\mbf{G}}+\wt{\sfn{D}}\det {\mbf{G}}\\&+\wh{\mbf{B}}^{\top}\cdot {\mbf{F}\mbf{G}^{-1}}(\det \mbf{G})+\wh{\mbf{C}}\cdot{{\mathrm{cof~}}}({\mbf{F}}{\mbf{G}}^{-1})(\det \mbf{G})+J_{\alpha j C}(.)_{\alpha j C})\Omega_v,\end{aligned}\end{equation}
where $(.)_{\alpha j C}=G_{\alpha A}F_{jB}\upepsilon_{CAB}=\mbf{G}^{\top}\mbs{e}_\alpha{\wedge}\mbf{F}^{\top}\mbs{e}_j\cdot\mbs{e}_C, (.)=\mbs{e}_\alpha\otimes \mbs{e}_j \otimes(\mbf{G}^{\top}\mbs{e}_\alpha{\wedge}\mbf{F}^{\top}\mbs{e}_j).$
Replacing the inverse of $\mbf{G}$ by the cofactor, we get the form \eqref{omegafn} which does not depend on the invertibility of $\mbf{G}.$
The components of the cofactor of $\mbf{A}$ are given by
\begin{equation}
\label{cofAcomp}
({\mathrm{cof~}} \mbf{A})_{ij}=\frac{1}{2}\upepsilon_{imn}\upepsilon_{jpq}\mbsi{A}_{mp}\mbsi{A}_{nq},
\end{equation}

\section{Expansion of $d\zeta$}
\label{appB}
The expression \eqref{zeta3d} leads to its exterior derivative 
\begin{equation}\begin{aligned}
d\zeta 
&=\frac{1}{2}\upepsilon_{ABC}d\mbsi{L}_A{{\wedge}} dx_B{{\wedge}} dx_C+d\mbsi{K}_{iA}{{\wedge}} dy_i {{\wedge}} dx_A+\frac{1}{2}\upepsilon_{ijk}d\mbsi{M}_i {{\wedge}} dy_j {{\wedge}} dy_k\\
&+d\wt{\mbsi{K}}_{\alpha A}{\wedge} d\theta_\alpha{{\wedge}} dx_A
+\frac{1}{2}\upepsilon_{\alpha\beta\gamma}d\wt{\mbsi{M}}_\alpha{\wedge} d\theta_\beta{{\wedge}} d\theta_\gamma
+d\mbsi{H}_{\alpha j}{\wedge} d\theta_\alpha{{\wedge}} dy_j,
\end{aligned}\end{equation}
which can be expanded further given that $\mbs{L}, \mbs{M}, \wt{\mbs{M}}, \mbf{K}, \wt{\mbf{K}}, \mbf{H}$ are functions of $(\mbs{x},\mbs{\chi},\mbs{\theta})$ so that
\begin{equation}\begin{aligned}
d\zeta 
&=\frac{1}{2}\upepsilon_{ABC}\mbsi{L}_{A,D}dx_D {{\wedge}} dx_B{{\wedge}} dx_C+\frac{1}{2}\upepsilon_{ABC}\mbsi{L}_{A,i}dy_i {{\wedge}} dx_B{{\wedge}} dx_C+\frac{1}{2}\upepsilon_{ABC}\mbsi{L}_{A,\alpha}d\theta_\alpha {{\wedge}} dx_B{{\wedge}} dx_C\\ 
&+\mbsi{K}_{iA,B}dx_B{{\wedge}} dy_i {{\wedge}} dx_A+\mbsi{K}_{iA,j}dy_j{{\wedge}} dy_i {{\wedge}} dx_A+\mbsi{K}_{iA,\beta}d\theta_\beta{{\wedge}} dy_i {{\wedge}} dx_A \\
&+\frac{1}{2}\upepsilon_{ijk}\mbsi{M}_{i,A}dx_A {{\wedge}} dy_j {{\wedge}} dy_k+\frac{1}{2}\upepsilon_{ijk}\mbsi{M}_{i,l}dy_l {{\wedge}} dy_j {{\wedge}} dy_k+\frac{1}{2}\upepsilon_{ijk}\mbsi{M}_{i,\alpha}d\theta_\alpha {{\wedge}} dy_j {{\wedge}} dy_k\\
&+\wt{\mbsi{K}}_{\alpha A,B}dx_B{{\wedge}} d\theta_\alpha {{\wedge}} dx_A+\wt{\mbsi{K}}_{\alpha A,j}dy_j{{\wedge}} d\theta_\alpha {{\wedge}} dx_A+\wt{\mbsi{K}}_{\alpha A,\beta}d\theta_\beta{{\wedge}} d\theta_\alpha {{\wedge}} dx_A \\
&+\frac{1}{2}\upepsilon_{\alpha\beta\gamma}\wt{\mbsi{M}}_{\alpha,A}dx_A {{\wedge}} d\theta_\beta {{\wedge}} d\theta_\gamma+\frac{1}{2}\upepsilon_{\alpha\beta\gamma}\wt{\mbsi{M}}_{\alpha,l}dy_l {{\wedge}} d\theta_\beta {{\wedge}} d\theta_\gamma+\frac{1}{2}\upepsilon_{\alpha\beta\gamma}\wt{\mbsi{M}}_{\alpha,\delta}d\theta_\delta {{\wedge}} d\theta_\beta {{\wedge}} d\theta_\gamma\\
&+\mbsi{H}_{\alpha j,A}dx_A{\wedge} d\theta_\alpha{{\wedge}} dy_j+\mbsi{H}_{\alpha j,k}dy_k{\wedge} d\theta_\alpha{{\wedge}} dy_j+\mbsi{H}_{\alpha j,\beta}d\theta_\beta{\wedge} d\theta_\alpha{{\wedge}} dy_j.
\end{aligned}\end{equation}
Collecting the terms accompanying the same exterior product of differentials, we get
\begin{equation}\begin{aligned}
d\zeta 
&=\mbsi{L}_{A,A}\Omega_v+(\frac{1}{2}\upepsilon_{ABC}\mbsi{L}_{A,i}- \mbsi{K}_{iC,B})dy_i {{\wedge}}dx_B {{\wedge}} dx_C\\
&+(\mbsi{K}_{kA,j}+\frac{1}{2}\upepsilon_{ijk}\mbsi{M}_{i,A})dy_j {{\wedge}} dy_k{\wedge} dx_A+\mbsi{M}_{i,i}\Omega_y+\wt{\mbsi{M}}_{\alpha,\alpha}\Omega_\theta\\
&+(\frac{1}{2}\upepsilon_{ijk}\mbsi{M}_{i,\alpha}+\mbsi{H}_{\alpha j,k})dy_j{\wedge} dy_k{\wedge} d\theta_\alpha+(\frac{1}{2}\upepsilon_{ABC}\mbsi{L}_{A,\alpha}-\wt{\mbsi{K}}_{\alpha C,B})d\theta_\alpha {{\wedge}} dx_B{{\wedge}} dx_C\\
&+(\wt{\mbsi{K}}_{\gamma A,\beta}+\frac{1}{2}\upepsilon_{\alpha\beta\gamma}\wt{\mbsi{M}}_{\alpha,A})d\theta_\beta {{\wedge}} d\theta_\gamma{\wedge} dx_A+(\mbsi{K}_{jC,\alpha}-\wt{\mbsi{K}}_{\alpha C,j}+\mbsi{H}_{\alpha j,C})d\theta_\alpha{{\wedge}} dy_j{\wedge} dx_C\\
&+(\frac{1}{2}\upepsilon_{\alpha\beta\gamma}\wt{\mbsi{M}}_{\alpha,i}+\mbsi{H}_{\gamma i,\beta})dy_i{{\wedge}} d\theta_\beta{\wedge} d\theta_\gamma,
\end{aligned}\end{equation}
where \eqref{omgv} is used.

\section{Expansion of ${\nabla\cdot}\mbs{P}$}
\label{appC}
The expression \eqref{P3d} is equivalent to
\begin{equation}\begin{aligned}
\mbs{P}
&= \mbs{L}+({\mbf{F}}^{\top}\mbf{K})^{\times}+({{\mathrm{cof~}}}{\mbf{F}})^{\top}\mbs{M}+({\mbf{G}}^{\top}\wt{\mbf{K}})^{\times}\\
&+({{\mathrm{cof~}}}{\mbf{G}})^{\top}\wt{\mbs{M}}+\upepsilon_{ABC}G_{\alpha B}F_{iC}\mbsi{H}_{\alpha i}\mbs{e}_A,
\end{aligned}\end{equation}
as
\begin{equation}\begin{aligned}
\upepsilon_{ABC}G_{\alpha B}F_{iC}\mbsi{H}_{\alpha i}\mbs{e}_A
&=\mbs{e}_A(\mbs{e}_A\cdot(\mbs{e}_B{\wedge}\mbs{e}_C))G_{\alpha B}F_{iC}\mbsi{H}_{\alpha i}\\
&=\mbs{e}_A\otimes\mbs{e}_A(\mbs{e}_B{\wedge}\mbs{e}_C)G_{\alpha B}F_{iC}\mbsi{H}_{\alpha i}\\
&=(G_{\alpha B}\mbs{e}_B{\wedge} F_{iC}\mbs{e}_C)\mbsi{H}_{\alpha i}\\
&=(\mbf{G}^{\top}\mbs{e}_\alpha{\wedge} \mbf{F}^{\top}\mbs{e}_j)\mbsi{H}_{\alpha j}=(\mbf{G}^{\top}\mbs{e}_\alpha{\wedge} \mbf{F}^{\top}\mbs{e}_j)\mbf{H}\cdot\mbs{e}_\alpha\otimes \mbs{e}_j\\
&=(\mbf{G}^{\top}\mbs{e}_\alpha{\wedge} \mbf{F}^{\top}\mbsi{H}_{\alpha j}\mbs{e}_j)=\mbf{G}^{\top}\mbs{e}_\alpha{\wedge} \mbf{F}^{\top}\mbf{H}^{\top}\mbs{e}_\alpha\\
&=\mathrm{axl}(\mbf{F}^{\top}\mbf{H}^{\top}\mbs{e}_\alpha\otimes\mbf{G}^{\top}\mbs{e}_\alpha -\mbf{G}^{\top}\mbs{e}_\alpha\otimes\mbf{F}^{\top}\mbf{H}^{\top}\mbs{e}_\alpha)\\
&=\mathrm{axl}(\mbf{F}^{\top}\mbf{H}^{\top}\mbf{G} -\mbf{G}^{\top}\mbf{H}\mbf{F})\\
&=-\frac{1}{2}(\mbf{F}^{\top}\mbf{H}^{\top}\mbf{G} -\mbf{G}^{\top}\mbf{H}\mbf{F})^{\times}
\end{aligned}\end{equation}
(recall $\mathrm{axl}(\mbs{b}\otimes\mbs{c}-\mbs{c}\otimes\mbs{b})=\mbs{c}{\wedge}\mbs{b}$).
Thus,
using the conditions stated in \S\ref{section3d},
\begin{equation}\begin{aligned}
{\nabla\cdot}\mbs{P}
&=\sfn{A}+({{\nabla_y}}\mbs{L})^{\top}\cdot {\mbf{F}}+({{\nabla_\theta}}\mbs{L})^{\top}\cdot {\mbf{G}}\\
&+({\mathrm{Curl}_x} \mbf{K}\cdot {\mbf{F}}-{{\mathrm{cof~}}}{\mbf{F}}\cdot({{\mathrm{Curl}_y}}\mbf{K}^{\top})^{\top})+\upepsilon_{ABC}{F}_{iB}\mbsi{K}_{iC,\alpha}G_{\alpha A}\\
&+{{\mathrm{cof~}}}{\mbf{F}}\cdot({\nabla_x}\mbs{M}+({\nabla_y}\mbs{M})\mbf{F}+({\nabla_\theta}\mbs{M})\mbf{G})\\
&+({\mathrm{Curl}_x} \wt{\mbf{K}}\cdot {\mbf{G}}-{{\mathrm{cof~}}}{\mbf{G}}\cdot({{\mathrm{Curl}_\theta}}\wt{\mbf{K}}^{\top})^{\top})+\upepsilon_{ABC}{G}_{\alpha B}\wt{\mbsi{K}}_{\alpha C,i}F_{iA}\\
&+({{\mathrm{cof~}}}{\mbf{G}})\cdot({\nabla_x}\wt{\mbs{M}}+({\nabla_y}\wt{\mbs{M}})\mbf{F}+({\nabla_\theta}\wt{\mbs{M}})\mbf{G})\\
&+\upepsilon_{ABC}G_{\alpha B}F_{iC}\mbsi{H}_{\alpha i,A}+\upepsilon_{ABC}G_{\alpha B}F_{iC}\mbsi{H}_{\alpha i,\beta}G_{\beta A}\\
&+\upepsilon_{ABC}G_{\alpha B}F_{iC}\mbsi{H}_{\alpha i,j}F_{jA},
\end{aligned}\end{equation}
i.e.,
\begin{equation}\begin{aligned}
{\nabla\cdot}\mbs{P}
& =\sfn{A}+\mbf{B} \cdot {\mbf{F}}+\mbf{C}\cdot{{\mathrm{cof~}}} {\mbf{F}}+\sfn{D}\det {\mbf{F}}+\wt{\mbf{B}} \cdot {\mbf{G}}+\wt{\mbf{C}}\cdot{{\mathrm{cof~}}} {\mbf{G}}+\wt{\sfn{D}}\det {\mbf{G}}\\
&+({{\mathrm{cof~}}}{\mbf{F}})\mbf{G}^{\top}\cdot({\nabla_\theta}\mbs{M})+({\nabla_y}\wt{\mbs{M}})^{\top}\cdot\mbf{F}({{\mathrm{cof~}}}{\mbf{G}})^{\top}\\
&+\upepsilon_{ABC}{F}_{iB}\mbsi{K}_{iC,\alpha}G_{\alpha A}+\upepsilon_{ABC}{G}_{\alpha B}\wt{\mbsi{K}}_{\alpha C,i}F_{iA}+\upepsilon_{ABC}G_{\alpha B}F_{iC}\mbsi{H}_{\alpha i,A}\\
&+\upepsilon_{ABC}G_{\beta A}G_{\alpha B}F_{iC}\mbsi{H}_{\alpha i,\beta}-\upepsilon_{ABC}F_{jA}F_{iB}G_{\alpha C}\mbsi{H}_{\alpha i,j},
\end{aligned}\end{equation}
i.e.,
\begin{equation}\begin{aligned}
{\nabla\cdot}\mbs{P}
& =\sfn{A}+\mbf{B} \cdot {\mbf{F}}+\mbf{C}\cdot{{\mathrm{cof~}}} {\mbf{F}}+\sfn{D}\det {\mbf{F}}+\wt{\mbf{B}} \cdot {\mbf{G}}+\wt{\mbf{C}}\cdot{{\mathrm{cof~}}} {\mbf{G}}+\wt{\sfn{D}}\det {\mbf{G}}\\
&+\wh{\mbf{B}}\cdot ({{\mathrm{cof~}}}\mbf{G})\mbf{F}^{\top}+\wh{\mbf{C}}\cdot({{\mathrm{cof~}}}{\mbf{F}}){\mbf{G}}^{\top}\\
&-({{\mathrm{cof~}}}{\mbf{F}})\mbf{G}^{\top}\cdot({\mathrm{Curl}_y}\mbf{H})^{\top}+({\mathrm{Curl}_\theta}\mbf{H}^{\top})\cdot\mbf{F}({{\mathrm{cof~}}}{\mbf{G}})^{\top}\\
&+\mathtt{J}\cdot\mbs{e}_\alpha\otimes \mbs{e}_j \otimes(\mbf{G}^{\top}\mbs{e}_\alpha{\wedge}\mbf{F}^{\top}\mbs{e}_j)\\
&+\upepsilon_{ABC}{F}_{iB}\mbsi{K}_{iC,\alpha}G_{\alpha A}+\upepsilon_{ABC}{G}_{\alpha B}\wt{\mbsi{K}}_{\alpha C,i}F_{iA}\\
&-\upepsilon_{ABC}(\mbsi{K}_{iC,\alpha}-\wt{\mbsi{K}}_{\alpha C,i}+\mbsi{H}_{\alpha i,C})G_{\alpha A}F_{iB}+\upepsilon_{CAB}G_{\alpha A}F_{iB}\mbsi{H}_{\alpha i,C}\\
&+\upepsilon_{ABC}G_{\alpha A}G_{\beta B}F_{iC}\mbsi{H}_{\beta i,\alpha}-\upepsilon_{ABC}F_{iA}F_{jB}G_{\alpha C}\mbsi{H}_{\alpha j,i},
\end{aligned}\end{equation}
so that finally,
\begin{equation}\begin{aligned}
{\nabla\cdot}\mbs{P}
&=\sfn{L}-({{\mathrm{cof~}}}{\mbf{F}})\mbf{G}^{\top}\cdot({\mathrm{Curl}_y}\mbf{H})^{\top}+({\mathrm{Curl}_\theta}\mbf{H}^{\top})\cdot\mbf{F}({{\mathrm{cof~}}}{\mbf{G}})^{\top}\\
&+\upepsilon_{\alpha\beta\gamma}({\mathrm{cof~}} \mbf{G})_{\gamma C}F_{iC}H^{\top}_{i\beta,\alpha}-\upepsilon_{ijk}({\mathrm{cof~}} \mbf{F})_{kC}G_{\alpha C}\mbsi{H}_{\alpha j,i}\\
&=\sfn{L}.
\end{aligned}\end{equation}
Note that
\begin{equation}\begin{aligned}
\upepsilon_{ABC}\mbsi{A}_{i A}\mbsi{A}_{j B}B_{ij C}&=
\upepsilon_{ABC}\mbsi{A}_{i A}\mbsi{A}_{j B}\delta_{CD}B_{ij D}\\
&=\upepsilon_{ABC}\mbsi{A}_{i A}\mbsi{A}_{j B}A^{\top}_{Ck }(\mbf{A}^{-\top})_{kD}B_{ij D}\\
&=\upepsilon_{ijk}({\mathrm{cof~}} \mbf{A})_{kD}{B}_{ijD}.
\end{aligned}\end{equation}

\end{document}